\documentclass[12pt]{article}

\usepackage{amsfonts, amsmath, amssymb, bm, bbm, bbold, latexsym}
\usepackage{esint} 
\usepackage{pstricks}
\usepackage{pst-node}
\usepackage{epsfig}
\usepackage{xcolor}
\usepackage{hyperref}
\usepackage{slashed} 

\newcommand{\orderof}{\mathcal{O}}
\newcommand{\half}{\frac{1}{2}}
\newcommand{\beq}{\begin{equation}}
\newcommand{\eeq}{\end{equation}}
\newcommand{\bea}{\begin{eqnarray}}
\newcommand{\eea}{\end{eqnarray}}

\newcommand{\dif}{\mathrm{d}}

\usepackage{import}

\usepackage{enumerate}

\usepackage{physics}

\usepackage[lmargin=1.0in, rmargin=1.0in, tmargin=1.0in, bmargin=1.0in]{geometry}

\usepackage{cancel}

\usepackage{empheq}

\usepackage{comment}

\usepackage{setspace}
\begin{document}



\title{\bf Symmetric teleparallel gravitational effects on solar neutrino oscillations}
\author{Aysel Cetinkaya$^1$, Muzaffer Adak$^1$, Ozcan Sert$^1$, Caglar Pala$^2$  \\
   {\small $^1$Computational and Gravitational Physics Laboratory, Department of Physics,} \\
   {\small Faculty of Sciences, Pamukkale University, Denizli, Türkiye} \\
  {\small $^2$Department of Physics,
Faculty of Arts and Sciences, Erciyes University, Kayseri, Türkiye} \\
      {\small {\it E-mail:} {ayselcetnkya@gmail.com, madak@pau.edu.tr, osert@pau.edu.tr, caglar.pala@gmail.com}}}

  \vskip 1cm
\date{\today}
\maketitle
\thispagestyle{empty}

\begin{abstract}

\noindent 
Neutrino oscillations probe the quantum gravity interface in unique ways. While gravitational effects on neutrinos are well studied in general relativity and torsion based geometries, the symmetric teleparallel regime where gravity stems solely from non-metricity, with zero curvature and torsion has remained uncharted. In this work, we perform the first analysis of neutrino oscillations in such a spacetime. Using the reduced Kerr metric in coincident gauge for the slowly rotating and weakly gravitating spherical Sun, we derive the Dirac Hamiltonian from the generalized Dirac equation and compute the accumulated phase of neutrino mass eigenstates. There are six free coupling constants in our model. Based on certain observational inputs, we inferred upper bounds on our arbitrary coupling constants. This allowed us to simplify the otherwise cumbersome calculations to some extent. Ultimately, we computed the phase differences that play a crucial role in solar neutrino oscillations and analyzed the contributions arising from our arbitrary coupling constants. Our results establish neutrino oscillations as a novel probe of non-metricity and open a new avenue for testing symmetric teleparallel gravity through astrophysical observations.\\
  

\noindent
{\it Keywords}: Metric-affine geometry, Symmetric teleparallel gravity, Neutrino oscillations, Non-metricity, Kerr metric, Solar neutrinos.

\end{abstract}

\noindent 
{\it Dedicated to Metin Gürses on the occasion of his becoming an emeritus professor after 60 years of research.}

\section{Introduction}

Neutrinos, the most elusive messengers of the cosmos, traverse astronomical distances while subtly encoding the structure of the spacetime through which they propagate. Their oscillatory behavior between flavor eigenstates not only resolved the long-standing solar neutrino puzzle but also opened an unexpected window onto physics beyond the Standard Model. Yet a deeper question remains largely unexamined: how does the geometry of spacetime itself shape these oscillations when gravity departs from its familiar Einsteinian form?

Neutrino oscillations, first proposed by Pontecorvo~\cite{Pontecorvo1958} and later confirmed by solar~\cite{SNO2001,SNO2002}, atmospheric~\cite{SuperK1998}, reactor~\cite{KamLAND2002,DayaBay2012}, and accelerator experiments~\cite{T2K2011,NOvA2016}, stand among the most profound discoveries in modern particle physics. The mere fact that neutrinos change flavour during propagation unequivocally demonstrates their non-zero masses and that flavour eigenstates are quantum superpositions of mass eigenstates—a phenomenon squarely outside the original Standard Model. In solar neutrino oscillations, a particle that essentially leaves the solar surface as an electron neutrino is detected at the Earth's surface as a muon neutrino. Therefore, the measured electron neutrino flux is smaller than expected.

The interplay between neutrino physics and gravity has attracted considerable attention, as neutrinos furnish unique probes of spacetime structure on scales inaccessible to other particles. Within general relativity (GR), gravitational effects on neutrino oscillations have been extensively studied~\cite{Cardall1996}-\cite{Swami2022}, revealing phase modifications through both gravitational redshift and geometric contributions.

Einstein’s GR is a mathematically coherent theory that has achieved remarkable success in accounting for a broad spectrum of observational phenomena. Within this framework, gravity is geometrically interpreted through the curvature derived from the metric structure of Riemannian spacetime. Nevertheless, the difficulty of explaining dark matter and dark energy within the theory, together with the absence of a fully consistent quantization and related conceptual challenges, provides compelling motivation to investigate alternative gravitational models \cite{L_Heisenberg2019}. Several avenues exist for modifying gravity, and generalized geometric frameworks extending beyond Riemannian spacetime present especially intriguing prospects.

In this work, we adopt the gravitational theory formulated in symmetric teleparallel spacetime ($R^a{}_b=0$, $T^a =0$, $Q_{ab} \neq 0$), namely symmetric teleparallel gravity (STPG) \cite{adak2005}-\cite{L_Heisenberg2023}. Previously, possible gravitational effects on solar neutrino oscillations were analyzed within Einstein–Cartan gravity ($R^a{}_b \neq 0$, $T^a \neq 0$, $Q_{ab} = 0$) \cite{Adak2001} and Einstein–Cartan–Weyl gravity ($R^a{}_b \neq 0$, $T^a \neq 0$, $Q_{ab} \neq 0$) \cite{Adak2004}, using a method based on computing the dynamical phase of massive neutrinos derived from the Dirac equation. Following a methodology parallel to those earlier studies, we here revisit solar neutrino oscillations in the framework of STPG. In doing so, we employ a recently generalized version of the Dirac equation \cite{Adak2025}. Compared to Riemann–Cartan and Riemann–Cartan–Weyl geometries, symmetric teleparallel geometry has the additional advantage of admitting a metric formulation in the coincident gauge. Despite the increasing theoretical attention devoted to STPG, its consequences for neutrino physics have so far remained completely unexplored. The present paper seeks to address this gap by providing the first systematic investigation of neutrino oscillations in symmetric teleparallel spacetime.

The mathematical foundations underlying our analysis including Clifford algebra conventions, spinor representations of the Lorentz group, covariant derivatives in metric-affine geometry and the construction of the generalized Dirac equation have been developed in detail in our previous work and related studies~\cite{Adak2025}. In the present paper, we do not repeat these derivations, but focus instead on their physical implications for neutrino oscillations in symmetric teleparallel spacetime. 

Our approach is as follows. We consider solar neutrinos propagating in the gravitational field of the Sun, modeled by the Kerr metric in the coincident gauge---a natural gauge choice in STPG---that simplifies the connection while preserving the physical content. We employ the generalized Dirac equation in metric-affine spacetime to derive the Hamiltonian governing mass neutrino evolution, and we calculate the dynamical phases acquired by mass eigenstates as they travel from the solar surface to the Earth's surface. Our main results can be summarized as follows:
\begin{enumerate}
    \item Non-metricity parameters $a_2$ and $a_4$ contribute to the oscillation phase, but \textit{only} when the two mass eigenstates have opposite spin orientations.
    \item The parameter $b_3$ induces a universal phase shift that affects all spin configurations.
    \item The Sun's angular momentum (rotation) does not affect the oscillation phase at first order in $\orderof{(a/r)}$.
    \item Consistency with solar neutrino data places upper bounds $|a_1| \sim |a_3| \lesssim 1$, $|a_2| \sim |a_4| \lesssim 10^3$ and $|b_3| \sim |b_4| \lesssim 10^{-33}$ on quantum gravity correction parameters.
\end{enumerate}

The paper is organized as follows. Section~\ref{sec:framework} reviews the differential geometric framework. Section~\ref{sec:dirac} establishes the generalized Dirac equation. Section~\ref{sec:hamiltonian} deals with neutrinos moving along the solar equatorial plane under the assumption of a weak gravitational field and slow rotational velocity. Section~\ref{sec:bounds} discusses observational constraints. Section~\ref{sec:discussion} presents our conclusions.

Throughout this paper, we use natural units $\hbar = c = G=1$ unless otherwise stated, and adopt the metric signature $(-,+,+,+)$. Greek indices $\mu, \nu, \ldots$ denote coordinate (holonomic) components, while Latin indices $a, b, \ldots$ denote Lorentz (anholonomic) components.
\section{Metric-affine spacetimes}
\label{sec:framework}

A metric-affine geometry is characterized by the triple $\{M, g, \nabla\}$ or equivalently by the set $\{Q_{ab}, T^a, R^a{}_b\}$, where $M$ is a four-dimensional orientable differentiable manifold, $g$ is a non-degenerate symmetric covariant metric tensor, $\nabla$ is the affine connection, $Q_{ab}$ is the non-metricity 1-form, $T^a$ is the torsion 2-form, and $R^a{}_b$ is the curvature 2-form \cite{thirring1997,frankel2012}. We denote the anholonomic metric-orthonormal coframe by $e^a$ and write the metric as $g = \eta_{ab}\, e^a \otimes e^b$, where $\eta_{ab} := \mathrm{diag}(-1,+1,+1,+1)$ is the Minkowski metric appropriate for (1+3)-dimensional spacetime.

The affine connection is determined by the full connection 1-form $\omega^a{}_b$ via the definition $\nabla e^a := -\omega^a{}_b \wedge e^b$, or equivalently $\nabla X_a := X_b \otimes \omega^b{}_a$, where $\otimes$ denotes the tensor product, $\wedge$ is the exterior product, and $X_a$ is the orthonormal frame dual to $e^a$ such that $e^a(X_b) = \delta^a_b$. The Cartan structure equations define the tensor-valued non-metricity 1-form, torsion 2-form, and curvature 2-form, respectively:
\begin{subequations}\label{eq:cartan}
 \begin{align}
    Q_{ab} &:= -\half D\eta_{ab} = \omega_{(ab)}\,, \label{eq:non-metricity}\\
    T^a &:= De^a = \dif e^a + \omega^a{}_b \wedge e^b\,, \label{eq:torsion}\\
    R^a{}_b &:= D\omega^a{}_b := \dif\omega^a{}_b + \omega^a{}_c \wedge \omega^c{}_b\,, \label{eq:curvature}
  \end{align}
\end{subequations}
where $\dif$ is the exterior derivative and $D$ is the covariant exterior derivative. Here we used $\dif\eta_{ab} = 0$ in the first equation. Throughout the paper, the notations $(\bullet)$ and $[\bullet]$ denote symmetrization and anti-symmetrization with respect to the enclosed indices: $\omega_{(ab)} = \half(\omega_{ab} + \omega_{ba})$ and $\omega_{[ab]} = \half(\omega_{ab} - \omega_{ba})$. The \textit{first kind trace} (Weyl) $Q$ and independently the \textit{second kind trace} $P$ 1-forms constructed from the non-metricity will be used in the forthcoming sections:
  \beq
    Q := \eta^{ab}Q_{ab} \qquad \text{and}  \qquad P := (\iota^a Q_{ab})e^b\,,
    \eeq
where $\iota_a := \iota_{X_a}$ denotes the interior product (or interior derivative) with respect to the orthonormal frame $X_a$. 

Geometry is classified according to whether non-metricity, torsion, and/or curvature vanish. Accordingly, symmetric teleparallel geometry is defined by the vanishing of both curvature and torsion, $R^a{}_b =0$ and $T^a=0$, while non-metricity remains unrestricted, $Q_{ab} \neq 0$. The condition $R^a{}_b = 0$ implies that the connection is flat, meaning that parallel transport is path-independent. Combined with $T^a = 0$, this ensures the existence of global inertial frames where the connection can be made to vanish in the holonomic coordinate bases—the \textit{natural gauge or inertial gauge or coincident gauge} \cite{adak2006,2018-koivisto}. In the orthonormal bases, the coincident gauged connection 1-form takes the simple form
\beq
\omega^a{}_b = h^a{}_\alpha \dif h^\alpha{}_b\,,
\label{eq:coincident}
\eeq
where $h^a{}_\alpha$ and $h^\alpha{}_a$ are the tetrad (vierbein) components relating orthonormal and coordinate bases:
\beq
e^a = h^a{}_\alpha \dif x^\alpha \qquad \text{and} \qquad \dif x^\alpha = h^\alpha{}_a e^a\,.
\label{eq:tetrad}
\eeq
They satisfy $h^a{}_\alpha h^\alpha{}_b = \delta^a_b$ and $h^\alpha{}_a h^a{}_\beta = \delta^\alpha_\beta$. This gauge choice drastically simplifies calculations while retaining the full non-metricity content of the geometry. 

\section{Dirac equation in metric-affine spacetimes}
\label{sec:dirac}

The propagation of spin-$\half$ fermions in metric-affine geometry requires specifying how the spinor covariant derivative couples to the connection. Recently, its generalized form was established in Ref.~\cite{Adak2025}. For the symmetric teleparallel geometric sector with vanishing torsion, the covariant exterior derivative of spinor takes the form
\beq
D\psi = \dif\psi + \Omega\psi\,,
\label{eq:Dpsi}
\eeq
where $\psi$ is a Dirac spinor and the spinor connection 1-form is
\bea
\Omega = \half\sigma_{ab}\omega^{[ab]} + (a_1 I + a_2\gamma_5)Q 
 + (a_3 I + a_4\gamma_5)P + (b_3\gamma_a + b_4\gamma_a\gamma_5)e^a\,.
\label{eq:Omega}
\eea
Here $\sigma_{ab} = \frac{1}{4}[\gamma_a,\gamma_b]$ are the Lorentz generators, $\gamma_a$ are the Dirac matrices satisfying $\{\gamma_a,\gamma_b\} = 2\eta_{ab}I$, $\gamma_5 = \gamma_0\gamma_1\gamma_2\gamma_3$, and ${a_1, a_2, a_3, a_4, b_3, b_4}$ are complex coupling constants.

In Riemannian spacetime, only one of the basis elements of the Clifford algebra, namely $\sigma_{ab}$, appears in the spinor connection. However, if one requires that all basis elements of the Clifford algebra be present in the spinor connection in a metric-affine spacetime, this combination plus the term $(b_1 I + b_2 \gamma_5)\, T$ emerge, where $T=\iota_a T^a$ is the torsion trace $1$-form, see \cite{Adak2025}. We suspect that the new coupling constants encode traces of novel types of interactions of the spinor. For the time being, the physical interpretation of these constants remains an open problem. The present work has been carried out precisely with this aim. Meanwhile, the coupling constants must satisfy specific constraints for consistency between the variational and canonical Dirac equations:
 \begin{subequations} \label{eq:constraints}
   \begin{align}
    a_1 &= -\half + iA_1\,, & a_2 &= A_2\,, & a_3 &= +\half + iA_3\,, \\
     a_4 &= A_4\,, &  b_3 &= B_3\,,  & b_4 &= B_4\,.
   \end{align}
  \end{subequations}
where $\{A_1, A_2, A_3, A_4, B_3, B_4\}$ are real constants and $i$ is the imaginary unit. Then, the generalized Dirac equation in differential form notation reads
\beq
\ast\gamma \wedge D\psi + m\psi\ast 1 = 0\,, \label{eq:Dirac}
\eeq
where $\gamma = \gamma_a e^a$ is the Clifford-algebra-valued 1-form, $\ast$ denotes the Hodge dual map, $m$ is the fermion mass, and $*1=e^0 \wedge e^1 \wedge e^2 \wedge e^3$ is the invariant volume 4-form. 

\section{Neutrinos propagating at the equator of the Sun}
\label{sec:hamiltonian}

Solar neutrinos propagate through the gravitational field of a slowly rotating spherical mass. The appropriate spacetime description is the Kerr metric in Boyer-Lindquist coordinates $(t,r,\theta,\phi)$. The rotational speed of the Sun about its own axis ($\sim 10^{3} \mathrm{m/s}$) is much smaller than the speed of light in a vacuum. Thus, the rotation parameter $a/r \ll 1$, permits a linearized treatment. Accordingly, in the equatorial plane ($\theta = \pi/2$), the reduced Kerr line element is
\beq
\dif s^2 \simeq -f^2 \dif t^2 + f^{-2}\dif r^2 + r^2\dif\phi^2 - \frac{4aM}{r}\dif t\dif\phi\,,
\label{eq:Kerr}
\eeq
where $f^2 \simeq 1 - 2M/r$, with the Sun mass $M$. The orthonormal coframe appropriate to this metric is
 \beq
    e^0 \simeq f \dif t - af \dif\phi \, , \quad e^1 \simeq f^{-1}\dif r\, , \quad e^2 \simeq 0\,,\quad e^3 \simeq -\frac{a}{r}\dif t + r\dif\phi\,.  \label{eq:coframe}
 \eeq 
Now, we can write down the tetrad and its inverse,
  \begin{align}
    h^a{}_\alpha \simeq \begin{pmatrix}
         f & 0 & 0 & -af \\
         0 & 1/f & 0 & 0 \\
         0 & 0 & 0 & 0 \\
         -a/r & 0 & 0 & r
     \end{pmatrix} \quad \text{and} \quad 
    h^\alpha{}_a  \simeq \begin{pmatrix}
         1/f & 0 & 0 & a/r \\
         0 & f & 0 & 0 \\
         0 & 0 & 0 & 0 \\
         a/fr^2 & 0 & 0 & 1/r
     \end{pmatrix} \, .
 \end{align} 
Using Eq.~\eqref{eq:coincident}, the coincident gauged orthonormal connection 1-form could be calculated,
   \begin{align}
     \omega^a{}_b  \simeq \begin{pmatrix}
         -f'e^1 & 0 & 0 & 0 \\
         0 & f'e^1 & 0 & 0 \\
         0 & 0 & 0 & 0 \\
         -2ae^1/r^2 & 0 & 0 & -fe^1/r 
     \end{pmatrix}
 \end{align}
where a prime denotes differentiation with respect to the radial coordinate $r$. Its symmetric piece yields the non-metricity 1-form:
   \beq
    Q_{ab} = \omega_{(ab)} \simeq  \begin{pmatrix}
    f' e^1 & 0 & 0 & -ae^1/r^2\\
    0 & f' e^1 & 0 & 0\\
    0 & 0 & 0 & 0\\
    -ae^1/r^2 & 0 & 0 & -f e^1/r 
    \end{pmatrix} \, .
   \label{eq:Qabmatrix}
   \eeq
Correspondingly, we can calculate the necessary quantities in the generalized Dirac equation Eq.~(\ref{eq:Dirac}).
   \begin{subequations}
    \begin{align}
     \omega_{[03]} &= ae^1/r^2 \, , \quad \text{other} \quad \omega_{[ab]} =0 \, ,  \\
     Q &= Q^a{}_a = \omega^a{}_a = -fe^1/r  \,, \label{eq:Qeval}\\
     P &= (\iota^a Q_{ab}) e^b  = f' e^1 \,. \label{eq:Peval}
\end{align}
\end{subequations}

Next, we substituted these results into equation (\ref{eq:Dirac}), computed the exterior derivative of the spinor, used the identity $*\gamma \wedge e^a = - \gamma^a *1$ together with the standard identities of the Dirac matrices, and wrote the components of the linear momentum operator in spherical coordinates as follows:
      \begin{align}
        p_r = -i \left( \frac{\partial}{\partial r} + \frac{1}{r} \right) \, , \qquad 
        p_\theta = - \frac{i}{r} \left( \frac{\partial}{\partial \theta} + \frac{\cot\theta}{2} \right) \, , \qquad p_\phi = - \frac{i}{r\sin\theta}  \frac{\partial}{\partial \phi} \, .
      \end{align}
Thus, for radially propagating neutrinos in the equatorial plane, i.e., $p_r=p$, $p_\theta =0$, $p_\phi=0$, we could express the generalized Dirac equation \eqref{eq:Dirac} in the Hamiltonian form:
\beq
i\frac{\partial\psi}{\partial t} = H\psi\,,
\label{eq:Schrodinger}
\eeq
where the spinor $\psi$ is a four-component column matrix and the Dirac Hamiltonian $H$ is a four-component square matrix. Moreover, let us recall that in the neutrino oscillation model the dynamical phases are determined by using the eigenvalues of the Hamiltonian, and the eigenfunction corresponding to a given eigenvalue is called a mass neutrino \cite{griffiths2008}. Accordingly, in the final form of the Dirac equation written in (\ref{eq:Dirac}), we employ the eigenvalue equation $H\psi = E \psi$. Let us denote the eigenvalues by the set, $ E = \{ E_+^\uparrow , E_+^\downarrow , E_-^\uparrow , E_-^\downarrow \}$. Thus, we assign and interpret the components of the eigenspinor as follows:
  \begin{align}
    \psi = \begin{bmatrix}
    \nu^\uparrow \\
    \nu^\downarrow \\
    \bar{\nu}^\uparrow \\
    \bar{\nu}^\downarrow \\
    \end{bmatrix} := \begin{bmatrix}
      \text{spin-up mass neutrino} \\
    \text{spin-down mass neutrino} \\
    \text{spin-up anti-mass neutrino} \\
    \text{spin-down anti-mass neutrino} \\
    \end{bmatrix} .
   \end{align}
As a result, the general Dirac equation takes the following form:
\begin{align}
i \frac{\partial \psi}{\partial t} = H \psi = E \psi \qquad \Rightarrow \qquad  \begin{bmatrix}
i\dot{\nu}^\uparrow \\
i\dot{\nu}^\downarrow \\
i\dot{\bar{\nu}}^\uparrow \\
i\dot{\bar{\nu}}^\downarrow \\
\end{bmatrix} = \begin{bmatrix}
E_+^\uparrow & 0 & 0 & 0 \\
0 & E_+^\downarrow & 0 & 0 \\
0 & 0 & E_-^\uparrow & 0 \\
0 & 0 & 0 & E_-^\downarrow \\
\end{bmatrix} \begin{bmatrix}
\nu^\uparrow \\
\nu^\downarrow \\
\bar{\nu}^\uparrow \\
\bar{\nu}^\downarrow \\
\end{bmatrix} ,
\end{align}
where a dot denotes the derivative with respect to time $t$. We are interested in neutrinos emitted from the Sun, rather than anti-neutrinos. For this reason, among the four scalar equations provided by this matrix equation, we shall proceed with the first two.
\begin{subequations}
\begin{align}
i \frac{\partial \nu^\uparrow}{\partial t} = E_+^\uparrow \nu^\uparrow \qquad \Rightarrow \qquad \nu^\uparrow (t) = \nu^\uparrow (0) e^{-i\Phi^\uparrow(t)} \\
i \frac{\partial \nu^\downarrow}{\partial t} = E_+^\downarrow \nu^\downarrow \qquad \Rightarrow \qquad \nu^\downarrow (t) = \nu^\downarrow (0) e^{-i\Phi^\downarrow(t)}
\end{align}
\end{subequations}
Here, the phases of the spin-up and spin-down mass neutrinos are given by
\begin{align} \label{eq:fazlar-salinim}
\Phi^\uparrow(t) = \int_0^{t} E_+^\uparrow (t) dt \qquad \text{and} \qquad \Phi^\downarrow(t) = \int_0^{t} E_+^\downarrow (t) dt
\end{align}
Above, $t=0$ denotes the production time of the neutrino on the solar surface, while $t > 0$ is the observation time on the Earth's surface; thus $\nu^\uparrow (0)$ and $\nu^\downarrow (0)$ are integration constants representing the initial state of the neutrinos. Since neutrinos are ultra-relativistic particles, in the phase integrals one may use $p \approx E$ and $dt \approx dr$:
\begin{align}
\Phi^\uparrow =  \int_{r_A}^{r_B} E_+^\uparrow (r) dr \qquad \text{and} \qquad \Phi^\downarrow = \int_{r_A}^{r_B} E_+^\downarrow (r) dr 
\end{align}
Here, since the origin of the coordinate system coincides with the center of the Sun, $r_A$ is the radial coordinate of the point where the solar electron neutrino is produced at time $t=0$ and may be taken approximately as the solar radius. The quantity $r_B$ is the radial coordinate of the point where the solar muon neutrino is detected at time $t>0$ and may be approximated by the distance between the center of the Sun and the surface of the Earth; hence $\Delta r = r_B - r_A$ is the distance between the solar surface and the Earth's surface.

In that case, what we need to do now is to compute the eigenvalues of the Dirac Hamiltonian in Eq.(\ref{eq:Schrodinger}) which is a complicated four-dimensional square matrix. We shall accomplish this in two steps. First, we bring the matrix $H$ into block-diagonal form; then, we diagonalize the upper block. For the first step, we perform the following calculation:
  \begin{subequations}
   \begin{align} \label{eq:block-diagonal}
      H\psi = H_{\pm} \psi \qquad \Rightarrow \qquad \text{det}(H-H_{\pm}I) =  \text{det}  \begin{pmatrix}
         H_{11}-H_{\pm} & H_{12} \\
         H_{21} & H_{22}-H_{\pm} \\
     \end{pmatrix} =0 \, ,\\
       H_{\pm} = \frac{1}{2} \left\{ \left( H_{11} + H_{22} \right) \pm \sqrt{\left( H_{11} + H_{22} \right)^2 -4 \left( H_{11} H_{22} - H_{12} H_{21} \right)}  \right\} \, ,
  \end{align}
  \end{subequations}
where $H_{11}, \, H_{12}, \, H_{21}, \, H_{22}$ are the two dimensional submatrices of $H$. Now, the two-dimensional square matrices $H_{\pm}$ are the diagonal elements of $H$. Then, under certain assumptions based on the observational data to be discussed later, $H_+$ has been approximately calculated as follows:
    \begin{align}
    H_+ \simeq  \begin{pmatrix}
         \mathcal{A} &  \mathcal{B} -i\mathcal{C} \\
          \mathcal{B} + i\mathcal{C} & \mathcal{A} \\ 
      \end{pmatrix} ,
  \end{align}
where 
   \begin{subequations} \label{eq:ozdeger-neut-osc1}
       \begin{align}
     \mathcal{A} &:=  f^2 p  + \frac{\mathcal{M}^2 + 16 b_4^2 }{2p} \, , \\
     \mathcal{B} &:=  a_4 f f' - \frac{a_2 f^2}{r} \, , \\
     \mathcal{C} &:= a \left( \frac{f}{2r^2} - i \frac{f^3}{r} p   + \frac{i \mathcal{M} f^2}{ r} \right) ,
 \end{align}
   \end{subequations}
where $\mathcal{M} := m - 4 b_3$. In the second step, we compute the eigenvalues of the Hamiltonian $H_+$ up to order $a$ by means of the eigenvalue equation $H_+ \psi_+ = E_+ \psi_+$:
       \begin{align} \label{eq:ozdeger-neut-osc2}
           E_+^\uparrow = \mathcal{A} + \mathcal{B} \qquad \text{and} \qquad  E_+^\downarrow = \mathcal{A} - \mathcal{B} \, .
       \end{align}

Finally, using the expressions (\ref{eq:ozdeger-neut-osc1}) and (\ref{eq:ozdeger-neut-osc2}) obtained above, we can explicitly evaluate the phase integrals given in (\ref{eq:fazlar-salinim}). At this stage, we introduce the assignment $\nu \to \nu_1$ to denote the first mass neutrino and $\nu \to \nu_2$ to denote the second mass neutrino. From this point onward, we explicitly restore all physical constants in their appropriate places.
      \begin{align}
     \Phi^\uparrow_1  &\approx  \frac{  m_1^2 c^4 - 8 b_3 m_p m_1 c^4 + 16 (b_3^2 + b_4^2) m_p^2 c^4 + 2E^2}{2\hbar c E}  \Delta r \nonumber \\ 
          & \qquad  + (2a_2 + a_4) \frac{MG \Delta r}{c^2 r_A r_B} - \left( a_2 + \frac{2MGE}{\hbar c^3} \right) \ln\frac{r_B}{r_A}  
 \end{align}
where we used $f\simeq 1-M/r$ for weak field approximation and $m_p$ is the Planck mass.  For the subsequent computation of $\Phi_1^\downarrow$, it is sufficient in the above result to make the substitutions $a_2 \to -a_2$ and $a_4 \to -a_4$.
   \begin{align}
     \Phi^\downarrow_1  &\approx  \frac{  m_1^2 c^4 - 8 b_3 m_p m_1 c^4 + 16 (b_3^2 + b_4^2) m_p^2 c^4 + 2E^2}{2\hbar c E}  \Delta r \nonumber \\ 
          & \qquad  - (2a_2 + a_4) \frac{MG \Delta r}{c^2 r_A r_B} + \left( a_2 - \frac{2MGE}{\hbar c^3} \right) \ln\frac{r_B}{r_A}  
 \end{align}
The same calculations and interpretations are repeated by replacing $\nu_1$ with $\nu_2$.
    \begin{subequations}
    \begin{align}
     \Phi^\uparrow_2  &\approx  \frac{  m_2^2 c^4 - 8 b_3 m_p m_2 c^4 + 16 (b_3^2 + b_4^2) m_p^2 c^4 + 2E^2}{2\hbar c E}  \Delta r \nonumber \\ 
          & \qquad  + (2a_2 + a_4) \frac{MG \Delta r}{c^2 r_A r_B} - \left( a_2 + \frac{2MGE}{\hbar c^3} \right) \ln\frac{r_B}{r_A}  \\
     \Phi^\downarrow_2  &\approx  \frac{  m_2^2 c^4 - 8 b_3 m_p m_2 c^4 + 16 (b_3^2 + b_4^2) m_p^2 c^4 + 2E^2}{2\hbar c E}  \Delta r \nonumber \\ 
          & \qquad  - (2a_2 + a_4) \frac{MG \Delta r}{c^2 r_A r_B} + \left( a_2 - \frac{2MGE}{\hbar c^3} \right) \ln\frac{r_B}{r_A}  
 \end{align}
   \end{subequations}
According to the solar neutrino oscillation model, the probability that a particle produced as an electron neutrino at the solar surface is detected as a muon neutrino at the Earth’s surface is given by
\begin{align} \label{eq:notr-oscil2}
P_{\nu_e \to \nu_\mu} &= \sin^2(2\theta) \, \sin^2\left( \frac{\Delta \Phi}{2} \right)
\end{align}
where $\Delta \Phi = \Phi_2 - \Phi_1$ is the phase difference of the mass neutrinos and $\theta$ is an empirical parameter, the so-called mixing angle \cite{griffiths2008}. Accordingly, there are four possible cases.
  \begin{itemize}
      \item Both mass eigenstates can be spin-up:
      \begin{align}
          \Delta \Phi = \Phi_2^\uparrow - \Phi_1^\uparrow = \frac{(\Delta m^2 - 8 b_3 m_p \Delta m) c^3 \Delta r}{2\hbar E} 
      \end{align}
       \item Both mass eigenstates can be spin-down:
      \begin{align}
          \Delta \Phi = \Phi_2^\downarrow - \Phi_1^\downarrow = \frac{(\Delta m^2 - 8 b_3 m_p \Delta m) c^3 \Delta r}{2\hbar E} 
      \end{align}
       \item $\nu_2$ can be spin-up while $\nu_1$ can be spin-down:
      \begin{align}
          \Delta \Phi &= \Phi_2^\uparrow - \Phi_1^\downarrow = \frac{(\Delta m^2 - 8 b_3 m_p \Delta m) c^3 \Delta r}{2\hbar E} \nonumber \\ 
          & \qquad  + 2(2a_2 + a_4) \frac{MG \Delta r}{c^2 r_A r_B} - 2\left( a_2 + \frac{2MGE}{\hbar c^3} \right) \ln\frac{r_B}{r_A} 
      \end{align}
       \item $\nu_2$ can be spin-down while $\nu_1$ can be spin-up:
      \begin{align}
          \Delta \Phi &= \Phi_2^\downarrow - \Phi_1^\uparrow = \frac{(\Delta m^2 - 8 b_3 m_p \Delta m) c^3 \Delta r}{2\hbar E} \nonumber \\ 
          & \qquad  - 2(2a_2 + a_4) \frac{MG \Delta r}{c^2 r_A r_B} + 2\left( a_2 + \frac{2MGE}{\hbar c^3} \right) \ln\frac{r_B}{r_A} 
      \end{align}
  \end{itemize}
Above, $\Delta m^2 := m_2^2 - m_1^2$ is called the mass-squared difference, and $\Delta m := m_2 - m_1$ is called the mass difference. Several striking features emerge from the phase analysis:
   \begin{enumerate}
\item \textit{Absence of rotation effects:} Since the Sun's angular momentum parameter $a$ does not appear in any of the phase differences to order $\mathcal{O}(a/r)$, we conclude that solar rotation has no observable effect on neutrino oscillations at leading order.

\item \textit{Universal mass shift:} The parameter $b_3$ induces a shift in the effective mass-squared difference: $\Delta m^2 \to \Delta m^2 - 8b_3 m_p \Delta m$. This effect is present regardless of spin configuration and represents a genuine modification of the oscillation wavelength.

\item \textit{Spin-dependent geometric phase:} The parameters $a_2$ and $a_4$ contribute only when the mass eigenstates have \textit{opposite} helicities. This asymmetric contribution, positive for more configuration and negative for the other, averages to zero for unpolarized neutrino beams but could manifest in polarization-sensitive experiments.

\item \textit{Absence of $a_1$ and $a_3$ effects:} The imaginary parts of the non-metricity couplings ($A_1$, $A_3$) do not affect the oscillation phase for ultrarelativistic neutrinos, as they enter only through the radial momentum operator and are suppressed by factors of $\hbar/pr \ll 1$.

\item \textit{Absence of $b_4$ effect:} The coupling constant $b_4$, which manifests itself in the dynamical phase proportional to the square of the Planck mass, disappears from the physically meaningful phase difference.
\end{enumerate}

\section{Phenomenological bounds}
\label{sec:bounds}

Solar neutrino experiments have precisely determined the oscillation parameters \cite{PDG2024}-\cite{Bandyopadhyay2008}:
\begin{align}
\Delta m^2_{21} \approx 7.5 \times 10^{-5}\,\mathrm{eV}^2  \qquad \text{and} \qquad 
\sin^2\theta_{12} \approx 0.307 \pm 0.013\,.
\label{eq:oscparams}
\end{align}
The absence of observed anomalies beyond standard oscillation physics constrains deviations from the expected phase. Requiring the geometric correction to be smaller than experimental uncertainties: $|8b_3 m_p \Delta m| \lesssim \delta(\Delta m^2) \sim 10^{-6}\,\mathrm{eV}^2$, and using $\Delta m \sim 10^{-2}\,\mathrm{eV}$ and $m_p \sim 10^{19}\,\mathrm{GeV}$, we obtain $|b_3| \lesssim 10^{-33}$. Similarly, requiring the effective mass not to exceed the measured value: $16 b_4^2 m_p^2 \lesssim m^2 \sim 10^{-4}\,\mathrm{eV}^2$ yields $|b_4| \lesssim 10^{-33}$.

On the other hand, since the measured momenta of solar neutrinos are of the order of $(5\text{–}15),\mathrm{MeV}/c$, we anticipated the upper bound $|a_1| \approx |a_3| \lesssim 1$ and carried out our calculations accordingly.

These bounds are extraordinarily stringent, reflecting the extreme weakness of Planck-scale-suppressed corrections to neutrino propagation. The spin-dependent contributions from $a_2$ and $a_4$ are more difficult to constrain directly, as they require knowledge of neutrino helicity correlations. For the geometric terms:
\beq
2(2a_2 + a_4)\frac{MG\Delta r}{c^2 r_A r_B} \sim 10^{-5}(2a_2 + a_4)\,,
\label{eq:a2a4estimate}
\eeq
using $M \sim 10^3\,\mathrm{m}$, $\Delta r \sim 10^{11}\,\mathrm{m}$, and $r_A r_B \sim 10^{19}\,\mathrm{m}^2$.
For this to remain below the oscillation phase uncertainty ($\sim 10^{-2}$), we require
\beq
|2a_2 + a_4| \lesssim 10^3\,.
\label{eq:a2a4bound}
\eeq
This bound is considerably weaker than those on $a_1$, $a_3$, $b_3$ and $b_4$, suggesting that future polarized neutrino experiments could provide a more sensitive probe of these parameters.

\section{Discussions and conclusions}
\label{sec:discussion}

We have presented the first systematic investigation of neutrino oscillations in symmetric teleparallel gravity. By employing the generalized Dirac equation in metric-affine spacetime \cite{Adak2025} specialized to the symmetric teleparallel geometric sector, we derived the complete neutrino Hamiltonian in the coincident gauge for a slowly rotating and weakly gravitating source like the Sun. Our central results are:
 \begin{enumerate}
\item The non-metricity coupling constants $b_3$ and $b_4$ induce a universal shift in the effective neutrino mass-squared difference, constrained by solar neutrino data to $|b_3| \sim |b_4| \lesssim 10^{-33}$.

\item The parameters $a_2$ and $a_4$ generate spin-dependent geometric phases that contribute only when mass eigenstates possess opposite helicities. In the article \cite{Adak2001}, this novel effect attributed to the torsion sector of Riemann–Cartan spacetime is associated, in the symmetric teleparallel geometry, with only non-metricity.

\item Solar rotation does not affect neutrino oscillations at leading order in $a/r$, consistent with the smallness of relativistic frame-dragging effects at solar system scales.
\end{enumerate}

The extremely stringent bounds on $b_3$ and $b_4$ arise from the Planck-scale suppression inherent in the coupling structure. While these parameters appear unmeasurable with current technology, they serve as consistency checks: any theory predicting $|b_3|$ or $|b_4| \gtrsim 10^{-33}$ would be ruled out by existing solar neutrino data.

More intriguing are the helicity-dependent contributions from $a_2$ and $a_4$. These parameters could, in principle, be probed by experiments sensitive to neutrino polarization asymmetries. The development of polarized neutrino sources or detection techniques sensitive to helicity correlations would open a new window on Lorentz-violating physics in the gravitational sector.

Several extensions of this work merit investigation. The full Kerr geometry beyond the equatorial plane would introduce additional angular-dependent phases. Atmospheric and accelerator neutrinos, traversing Earth's gravitational field, would probe different geometric configurations. The extension to metric-affine theories with nonvanishing torsion would allow direct comparison with the Einstein-Cartan results of Refs.~\cite{Adak2001,Adak2004}.

Finally, we note that symmetric teleparallel gravity provides a theoretically distinct framework from torsion-based alternatives, with potentially different predictions for other spin-dependent phenomena. The program of testing gravitational physics through precision neutrino observations has only begun.

\section*{Acknowledgments}

This study was supported by Scientific and Technological Research Council of Türkiye (TUBITAK) under Grant Number 124F325. The authors thank TUBITAK for their support. The authors sincerely thank the anonymous referees for their thorough and attentive review of the manuscript, and for their insightful suggestions and constructive criticisms, which have significantly contributed to improving the clarity and overall quality of the paper.


\end{document}